\documentclass[aps,11pt,prd,longbibliography,notitlepage,nofootinbib]{revtex4-1}
\usepackage[most]{tcolorbox}
\usepackage{slashed}

\usepackage{tikz}
\usepackage{feynmp-auto}
\usepackage{amsmath}
\usepackage[utf8]{inputenc}
\usepackage[T1]{fontenc} 
\usepackage{array}
\usepackage{bbm}
\usetikzlibrary{decorations.pathreplacing}
\usepackage{soul}	
\usepackage{xcolor}

\usepackage{epsfig}
\usepackage{color}
\usepackage{slashed}
\usepackage{comment}
\usepackage{epstopdf}
\usepackage{xspace}
\usepackage{mathrsfs}
\usepackage[euler]{textgreek}
\usepackage{amsmath}
\usepackage{amsthm}
\usepackage{amsfonts}
\usepackage{amssymb}
\usepackage{graphicx}
\graphicspath{ {Figures/} }
\usepackage[font={small}]{caption}
\usepackage{subcaption}
\usepackage{float}
\usepackage{multirow}
\usepackage[hang, flushmargin,bottom]{footmisc} 
\usepackage{placeins}
\usepackage{enumitem}
\usepackage{slashed}
\usepackage{bbm}
\usepackage{appendix}
\usepackage{tabularx}
\usepackage{siunitx}

\usepackage{titlesec}
\titleformat{\chapter}[display]
  {\normalfont\LARGE\bfseries}
  {\chaptertitlename\ \thechapter}{5pt}{\LARGE}
  \titlespacing*{\chapter}{0pt}{-20pt}{35pt}
\usepackage{bigstrut}
\usepackage{pst-node}
\usepackage{pstricks}
\usepackage{physics}
\usepackage{mathtools}
\usepackage{setspace}
\usepackage{fancyhdr}
\usepackage[makeroom]{cancel}
\usepackage{feyn}
\newcommand{\be}{\begin{equation}}
\newcommand{\ee}{\end{equation}}
\newcommand{\bes}{\begin{equation*}}
\newcommand{\ees}{\end{equation*}}

\newcommand{\e}{\text{e}}

\usepackage{hyperref}
\hypersetup{%
  colorlinks = true,
  linkcolor  = blue,
  citecolor  = blue,
  urlcolor   = blue,
}
\usepackage{soul}
\usepackage{todonotes}
\usepackage{xpatch}
\makeatletter
\xpretocmd{\todo}{\@bsphack}{}{}
\xapptocmd{\todo}{\@esphack}{}{}
\makeatother

\newcommand{\beq}{\begin{equation}}
\newcommand{\eeq}{\end{equation}}

\newcommand{\SU}{\,{\rm SU}}
\newcommand{\U}{\,{\rm U}}

\definecolor{green}{HTML}{008000}
\definecolor{goldenrod}{HTML}{DAA520}
\definecolor{magenta}{HTML}{FF00FF}
\definecolor{silver}{HTML}{C0C0C0}
\definecolor{indigo}{HTML}{4B0082}
\definecolor{skyblue}{HTML}{87CEEB}
\definecolor{darkgoldenrod}{HTML}{B8860B}
\definecolor{orange}{HTML}{FFA500}
\definecolor{yellow}{HTML}{FFFF00}
\definecolor{saddlebrown}{HTML}{8B4513}
\definecolor{blue}{HTML}{0000FF}
\definecolor{turquoise}{HTML}{40E0D0}
\definecolor{yellow}{HTML}{FFFF00}
\definecolor{white}{HTML}{FFFFFF}
\definecolor{whitesmoke}{HTML}{F5F5F5}

\newcommand{\myComment}[1]{}

\usepackage{xparse}

\usetikzlibrary{decorations.pathmorphing}	
\usetikzlibrary{decorations.markings}
\tikzset{
    vector/.style={decorate, decoration={snake}, draw},
	provector/.style={decorate, decoration={snake,amplitude=2.5pt}, draw},
	antivector/.style={decorate, decoration={snake,amplitude=-2.5pt}, draw},
    fermion/.style={draw=black, postaction={decorate},
        decoration={markings,mark=at position .55 with {\arrow[draw=black]{>}}}},
    fermionr/.style={draw=black, postaction={decorate},
    decoration={markings,mark=at position .55 with {\arrow[draw=black]{<}}}},
    fermioncyan/.style={draw=black, postaction={decorate},
        decoration={markings,mark=at position .55 with {\arrow[draw=cyan]{<}}}},
    fermiondif/.style={draw=black, postaction={decorate},
        decoration={markings,mark=at position .7 with {\arrow[draw=black]{>}}}},
            fermiondif2/.style={draw=black, postaction={decorate},
        decoration={markings,mark=at position .7 with {\arrow[draw=black]{<}}}},
    fermionend/.style={draw=black, postaction={decorate},
        decoration={markings,mark=at position 1 with {\arrow[draw=black]{>}}}},
    fermionuchannel2/.style={draw=black, postaction={decorate},
        decoration={markings,mark=at position .4 with {\arrow[draw=black]{>}}}},
    scalardif/.style={dashed,draw=black, postaction={decorate},
        decoration={markings,mark=at position .7 with {\arrow[draw=black]{>}}}},
    scalarend/.style={dashed,draw=black, postaction={decorate},
        decoration={markings,mark=at position 1 with {\arrow[draw=black]{>}}}},
    fermionbar/.style={draw=black, postaction={decorate},
        decoration={markings,mark=at position .55 with {\arrow[draw=black]{<}}}},
    fermionnoarrow/.style={draw=black},
    gluon/.style={decorate, draw=black,
        decoration={coil,amplitude=4pt, segment length=5pt}},
    scalar/.style={dashed,draw=black, postaction={decorate},
        decoration={markings,mark=at position .55 with {\arrow[draw=black]{>}}}},
    scalarcyan/.style={dashed,draw=black, postaction={decorate},
        decoration={markings,mark=at position .55 with {\arrow[draw=cyan]{>}}}},
    scalaruchannel1/.style={dashed,draw=black, postaction={decorate},
        decoration={markings,mark=at position .7 with {\arrow[draw=black]{>}}}},
                  scalaruchannel2/.style={dashed,draw=black, postaction={decorate},
        decoration={markings,mark=at position .4 with {\arrow[draw=black]{>}}}},
    scalarbar/.style={dashed,draw=black, postaction={decorate},
        decoration={markings,mark=at position .55 with {\arrow[draw=black]{<}}}},
    scalarnoarrow/.style={dashed,draw=black},
    electron/.style={draw=black, postaction={decorate},
        decoration={markings,mark=at position .55 with {\arrow[draw=black]{>}}}},
	bigvector/.style={decorate, decoration={snake,amplitude=4pt}, draw},
}

\NewDocumentCommand\semiloop{O{black}mmmO{}O{above}}
{%
\draw[#1] let \p1 = ($(#3)-(#2)$) in (#3) arc (#4:({#4+180}):({0.5*veclen(\x1,\y1)})node[midway, #6] {#5};)
}

\tikzstyle{block} = [draw, rectangle, 
    minimum height=3em, minimum width=6em]

\usepackage{enumitem}
\usepackage{tikz}
\usetikzlibrary{fit}
\tikzset{%
  highlight/.style={rectangle,rounded corners,color=granate,draw,text opacity =1,
    fill opacity=0.5,thick,inner sep=0pt}
}

\usepackage{xparse}
\NewDocumentCommand\loopv{O{black}mmmO{}O{above}}
{%
\draw[#1] let \p1 = ($(#3)-(#2)$) in (#3) arc (#4:({#4+360}):({0.5*veclen(\x1,\y1)})node[midway, #6] {#5};)
}

\usepackage{placeins}

\tikzset{
    cross/.pic = {
    \draw[rotate = 45] (-#1,0) -- (#1,0);
    \draw[rotate = 45] (0,-#1) -- (0, #1);
    }
}

\tikzset{
    square/.style={%
        draw=none,
        circle,
        append after command={%
            \pgfextra \draw[#1] (\tikzlastnode.north-|\tikzlastnode.west) rectangle 
                (\tikzlastnode.south-|\tikzlastnode.east);\endpgfextra}
    },
    square/.default=black
}

\tikzstyle{block} = [draw, rectangle, 
    minimum height=3em, minimum width=6em]
\begin{document}
\title{\LARGE{\textsf{Matter Unification and Lepton Flavour Violation}}}
\author{Hridoy Debnath, Pavel Fileviez P{\'e}rez}
\affiliation{
Physics Department and Center for Education and Research in Cosmology and Astrophysics, Case Western Reserve University, Cleveland, OH 44106, USA}
\email{hxd253@case.edu, pxf112@case.edu}
\date{\today}
\begin{abstract}
We explore the idea of quark–lepton unification at low energies. In particular, we discuss the minimal framework for matter unification at the multi-TeV scale, in which neutrino masses are necessarily generated via the inverse seesaw mechanism. To assess the testability of this theory for physics beyond the Standard Model, we analyze current experimental constraints and derive the corresponding lower bound on the symmetry breaking scale. We reexamine the impact of existing limits from lepton number violating meson decays, taking into account the freedom associated with unknown quark–lepton mixing angles. Furthermore, we study the correlation between bounds from meson decays and $\mu \to e$ conversion. We demonstrate that the upcoming $\mu \to e$ conversion experiment at Fermilab can play a crucial role in probing quark–lepton unification at the multi-TeV scale.
\end{abstract}
\maketitle
\section{INTRODUCTION}
The Standard Model (SM) of particle physics is one of the most successful theories for fundamental interactions. Nevertheless, it must be extended to account for the origin of neutrino masses and the observed matter–antimatter asymmetry of the Universe.
The idea of unification is among the most compelling frameworks for physics beyond the SM. In unified theories, neutrino masses can naturally arise via the seesaw mechanism, while the baryon asymmetry can be explained through leptogenesis.
However, conventional unification is realized at an extremely high energy scale, 
$M_{GUT} \sim 10^{15-16}$ GeV, far beyond the reach of current or foreseeable collider experiments, making direct experimental tests impossible. 

The idea of quark–lepton unification, originally proposed by J.~Pati and A.~Salam in Refs.~\cite{Pati:1973rp,Pati:1974yy}, provides an elegant theoretical framework for physics beyond the Standard Model. In this framework, quarks and leptons are placed within the same multiplets, and the gauge symmetry $SU(3)_C$ is unified with 
$U(1)_{B-L}$ into a larger $SU(4)_C$ gauge group, where $B$ and $L$ denote baryon and lepton number, respectively. A remarkable feature of this theory is that it predicted massive neutrinos well before their experimental confirmation. However, in its minimal realization, the model implies that neutrinos have masses comparable to those of the up-type quarks. Small neutrino masses can be obtained by implementing the seesaw mechanism~\cite{Minkowski:1977sc,Gell-Mann:1979vob,Mohapatra:1979ia,Yanagida:1979as}, but this typically requires the theory to be realized at the canonical seesaw scale, $M_{R} \sim 10^{14}$ GeV.

The possibility of achieving low-scale matter unification was investigated in Ref.~\cite{FileviezPerez:2013zmv}. In this minimal framework, the gauge symmetry is the simplest one required for matter unification, namely $SU(4)_C \otimes SU(2)_L \otimes U(1)_R$. To obtain a realistic theory, neutrino masses are generated via the inverse seesaw mechanism.
However, the symmetry-breaking scale is constrained from below by experimental limits on rare meson decays. In the simplest scenarios, these bounds require the $SU(4)_C$ breaking scale to lie above $10^3$ TeV. As we will discuss in detail, this lower bound can be relaxed once one accounts for the freedom associated with the presently unknown quark–lepton mixing angles. This minimal theory for matter unification has been studied in detail in Refs.~\cite{Butterworth:2025ttw,FileviezPerez:2023rxn,FileviezPerez:2022fni,FileviezPerez:2022rbk,FileviezPerez:2021arx,FileviezPerez:2021lkq,Flacke:2025xwl,Gedeonova:2022iac,Murgui:2021bdy}.

In this article, we perform a detailed study of the predictions for lepton flavor–violating (LFV) meson decays in order to determine the lower bound on the symmetry-breaking scale. We analyze the most general effective interactions relevant for meson decays, fully accounting for the unknown quark–lepton mixing angles.
In particular, we show that the effective Hamiltonian for 
$K_L^0\to e^\mp \mu^\pm$ mediated by vector leptoquarks contains eight independent Wilson coefficients, each determined by the unknown quark–lepton mixing parameters. We examine in detail how current experimental limits on LFV meson decays constrain the symmetry breaking scale. To illustrate the dependence on the mixing structure, we present four simple benchmark scenarios that clearly exhibit how the decay rates vary with the mixing angles.
In three of these cases, the LFV meson decays are highly suppressed and vanish at tree level. Such processes have been studied previously in Refs.~\cite{Valencia:1994cj,Gedeonova:2022iac,FileviezPerez:2022rbk,FileviezPerez:2021lkq}. However, our analysis is more general. By considering the full flavor structure of the theory, we derive robust lower bounds on the symmetry-breaking scale.

The new generation of charged lepton flavor violation experiments has the potential to probe physics at the multi-TeV scale~\cite{Bernstein:2013hba,Davidson:2022jai,Davidson:2022nnl,FileviezPerez:2022ypk}. In particular, the Mu2e experiment~\cite{Mu2e:2014fns} at Fermilab is expected to improve the experimental sensitivity to $\mu$ to $e$
conversion by several orders of magnitude. In this work, we present a detailed analysis of the tree-level contributions to $\mu$ to $e$ conversion mediated by vector and scalar leptoquarks. We show that the current experimental bounds 
in gold provide the most stringent constraints in scenarios where LFV meson decays are suppressed. We further explore the correlation between LFV meson decays and 
$\mu$ to $e$ conversion, demonstrating that the projected sensitivity of the Mu2e experiment will surpass the constraints from LFV meson decays and probe symmetry breaking scales as high as 
$10^4$ TeV.
These results highlight the crucial role of the Mu2e experiment in testing one of the most compelling frameworks for physics beyond the Standard Model—namely, the unification of matter.

This article is organized as follows. In Section~\ref{unification}, we briefly review the minimal framework for quark–lepton unification at low scales. Section~\ref{mesons} presents a detailed analysis of lepton flavor violating Kaon decays. In Section~\ref{mutoe}, we investigate the predictions for $\mu \to e$ conversion mediated by vector leptoquarks, while the corresponding contributions from scalar leptoquarks are examined in Section~\ref{scalarLQ}. Our main findings are summarized in Section~\ref{summary}.
%
\section{MATTER UNIFICATION}
\label{unification}
The minimal theory for matter unification is based on the gauge symmetry~\cite{FileviezPerez:2013zmv} $$\SU(4)_C \otimes \SU(2)_L \otimes \U(1)_R,$$ while the SM
matter fields are unified as follows
\begin{eqnarray}
F_{q_L} &=&
\left(
\begin{array}{cccc}
u_r & u_g & u_b  & \nu 
\\
d_r & d_g & d_b  & e
\end{array}
\right)_L \sim (\mathbf{4}, \mathbf{2}, 0), \\[1ex]
F_{u_R} &=&
\left(
\begin{array}{cccc}
u_r & u_g & u_b & \nu
\end{array}
\right)_R \sim (\mathbf{{4}}, \mathbf{1}, 1/2), 
\\[1ex]
 F_{d_R} &=&
\left(
\begin{array}{cccc}
d_r & d_g & d_b & e
\end{array}
\right)_R \sim (\mathbf{{4}}, \mathbf{1}, -1/2).
\end{eqnarray}
Here $u_i$ and $d_i$ (with $i=r,g,b$) are the quarks with different colors. The kinetic terms for our fermionic fields are given by
\begin{eqnarray}
\mathcal{L} &\supset&   i \bar{F}_{q_L} \slashed{D} F_{q_L} +  i \bar{F}_{u_R} \slashed{D} F_{u_R} +  i \bar{F}_{d_R} \slashed{D} F_{d_R}.
\end{eqnarray}
In our notation, the $\SU(4)_C$ gauge fields are, $V_\mu \sim (\mathbf{15},\mathbf{1},0)$,  the $\SU(2)_L$ gauge field is, $W_\mu \sim (\mathbf{1},\mathbf{3},0)$, while the $\U(1)_R$ gauge field is, $B_{R\mu} \sim (\mathbf{1},\mathbf{1},0)$.
The covariant derivatives for the 
fermionic fields are written as
\begin{eqnarray}
\slashed{D} F_{q_L} &=& \gamma^\mu (\partial_\mu + i g_4 V_\mu + i g_2 W_\mu) F_{q_L}, \\
\slashed{D} F_{u_R} &=&  \gamma^\mu (\partial_\mu + i g_4 V_\mu + \frac{i}{2}  g_R B_{R \mu} ) F_{u_R},   \\
\slashed{D} F_{d_R} &=&  \gamma^\mu (\partial_\mu + i g_4 V_\mu - \frac{i}{2}  g_R B_{R\mu} ) F_{d_R},
\end{eqnarray}
while the Yukawa interactions are given by
\begin{align}
	-\mathcal{L}_Y = Y_1 \bar{F}_{q_L} i \sigma_2 H_1^* F_{u_R} 
    + Y_2 \bar{F}_{u_R} \Phi^\alpha F_{q_L}^\beta \epsilon_{\alpha \beta} + Y_3 \bar{F}_{q_L} H_1 F_{d_R} + Y_4 \bar{F}_{q_L} \Phi F_{d_R} +  {\rm H.c.},
    \label{Yukawa}
\end{align}
with $H_1 \sim (\mathbf{1},\mathbf{2},1/2)$ and $\Phi \sim (\mathbf{15},\mathbf{2},1/2)$. Notice that both Higgs fields are needed to generate realistic fermion masses in a consistent manner~\cite{FileviezPerez:2013zmv}. In the above equation, $\alpha$ and $\beta$ are $SU(2)_L$ indices.
The $SU(4)_C$ gauge fields can be written as
  \begin{eqnarray}
    V^\mu =
    \left(
    \begin{array} {cc}
      G^\mu & X^\mu/\sqrt{2}  \\
      (X^\mu)^*/\sqrt{2} & 0  \\
    \end{array}
    \right) + T_{15} \ V_{15}^{\mu} \sim (\mathbf{15}, \mathbf{1},0),
  \end{eqnarray}
  where $G^\mu \sim (\mathbf{8},\mathbf{1},0)_\text{SM}$ are the SM gluons, $X^\mu \sim (\mathbf{3},\mathbf{1},2/3)_\text{SM}$ are vector leptoquark (LQ), and $V_{15}^{\mu} \sim (\mathbf{1},\mathbf{1},0)_\text{SM}$. Here $T_{15} = \rm{diag} (1,1,1,-3)/ 2 \sqrt{6}$. 
  Once the scalar field, $\chi = \left(  \chi_r  \  \chi_g \ \chi_b \ \chi^0 \right) \sim (\mathbf{4}, \mathbf{1}, 1/2)$, acquires a vacuum expectation value, $v_\chi$, the new massive vectors associated with the broken \mbox{generators} of $\SU(4)_C$ acquire mass. The mass of the vector leptoquarks, $X_\mu \sim ({\bf{3}},1,2/3)_{\rm SM}$, is given by
  \begin{equation}
    M_{X}^2   = \frac{1}{4} g_4^2 v_\chi^2\,.
  \end{equation}
This mass defines for us the $SU(4)_C$ symmetry breaking scale. In the next section, we will discuss the different ways to find the lower bound on the symmetry breaking scale using the experimental bounds from rare decays. This theory has been studied in detail in Refs.~\cite{Butterworth:2025ttw,FileviezPerez:2023rxn,FileviezPerez:2022fni,FileviezPerez:2022rbk,FileviezPerez:2021arx,FileviezPerez:2021lkq,Flacke:2025xwl,Gedeonova:2022iac}. It is very important to emphasize that this theory~\cite{FileviezPerez:2013zmv} can be realized at the multi-TeV scale thanks to the fact that Majorana neutrino masses are generated via the inverse seesaw mechanism~\cite{Mohapatra:1986bd}. This theory provides a concrete example for matter unification at the low scale and one can hope to test this idea at current or future collider experiments. A similar model can be found in Refs.~\cite{Valencia:1994cj,Smirnov:1995jq}, but without the needed mechanism for neutrino masses that allow us to have a realistic theory at the low scale. For other models for quark-lepton unification at the low scale, see for example Ref.~\cite{Babu:2025azv}.
  
Before we discuss the bounds from lepton number violating processes, we would like to mention that there are strong collider bounds on leptoquarks. When the vector leptoquark decays to $b\tau$ with $100\%$ branching ratio, the ATLAS collaboration excludes masses up to $1.9$ TeV~\cite{ATLAS:2023uox}. However, in the scenario where the leptoquark decays exclusively to a muon and a bottom quark pair, the CMS collaboration set a lower limit around $2$ TeV on the Leptoquark mass~\cite{CMS:2024bnj}. The non-resonant production of SM dileptons via t-channel LQ exchange provides the most stringent bound on the LQ mass. Depending on the coupling strength, the non-resonant dilepton production excludes vector LQ masses up to 5 TeV~\cite{CMS:2025iix,CMS:2022goy}. See also our recent article in Ref.~\cite{Butterworth:2025ttw} for a detailed study of collider bounds on scalar leptoquarks. 
%
\section{LEPTON FLAVOUR VIOLATING MESON DECAYS}
\label{mesons}
%
In order to find the lower bound on the $SU(4)_C$ breaking scale, we need to revisit the predictions for rare meson decays and use the current experimental bounds. The constraints from Kaon decays have been investigated from a long time, see, for example, the studies in Refs.~\cite{Valencia:1994cj,FileviezPerez:2022rbk}.
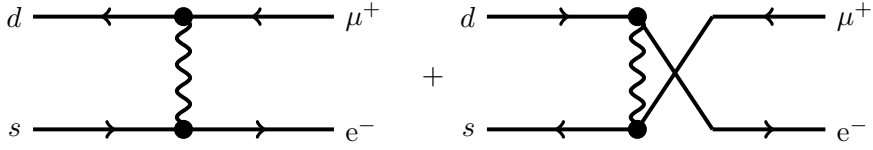
\begin{figure}[t]
\begin{eqnarray*}
\begin{gathered}
\begin{tikzpicture}[line width=1.5 pt,node distance=1 cm and 1.5 cm]
\coordinate[label =left: $d$] (i1);
\coordinate[right= 2cm of i1](v1);
\coordinate[below= 0.5cm of v1](vaux);
\coordinate[right = 2cm of v1, label= right:$\mu^+$](f1);
\coordinate[below = 1.5 cm of v1](v2);
\coordinate[left = 2 cm of v2, label=left: $s$] (i2);
\coordinate[right =  2 cm of v2,label=right: $\e^-$] (f2);
\draw[fermion] (v1) -- (i1);
\draw[vector] (v1) -- (v2);
\draw[fermion] (i2) -- (v2);
\draw[fermion] (v2) -- (f2);
\draw[fermion] (f1) -- (v1);
\draw[fill=black] (v1) circle (.1cm);
\draw[fill=black] (v2) circle (.1cm);
\end{tikzpicture}
\end{gathered}
\quad   \! +
\begin{gathered}
\begin{tikzpicture}[line width=1.5 pt,node distance=1 cm and 1.5 cm]
\coordinate[label =left: $d$] (i1);
\coordinate[right= 2cm of i1](v1);
\coordinate[below= 0.5cm of v1,label](vaux);
\coordinate[right = 1cm of v1](f1);
\coordinate[below = 1.5 cm of v1](v2);
\coordinate[left = 2 cm of v2, label=left: $s$] (i2);
\coordinate[right =  1 cm of v2] (f2);
\coordinate[right =  1.5 cm of f2,label=right: $\e^-$] (f3);
\coordinate[right =  1.5 cm of f1,label=right: $\mu^+$] (f4);
\draw[fermion] (i1) -- (v1);
\draw[vector] (v1) -- (v2);
\draw[fermion] (v2) -- (i2);
\draw[fermionnoarrow] (v1) -- (f2);
\draw[fermion] (f2) -- (f3);
\draw[fermion] (f4) -- (f1);
\draw[fermionnoarrow] (f1) -- (v2);
\draw[fill=black] (v1) circle (.1cm);
\draw[fill=black] (v2) circle (.1cm);
\end{tikzpicture}
\end{gathered}
\end{eqnarray*}
\caption{Feynman graphs for $K_L \to \mu^+ e^- $ decay. In the left-panel we show the contribution to $\bar{K}^0 \to e^- \mu^+$, while in the right-panel we show $K^0 \to e^- \mu^+$.}
\label{KLdecay}
\end{figure}

The relevant interactions for the vector leptoquark can be written as 
\begin{eqnarray}
    \bar{d}_i e_j X_\mu \hspace{0.3 cm } : \hspace{0.3 cm} i \frac{g_4}{\sqrt{2}} \gamma^\mu \left[V_L^{ij} P_L + V_R^{ij} P_R\right],
\end{eqnarray}
where $V_L=D_L^\dagger E_L$ and $V_R=D_R^\dagger E_R$ are the unknown mixing matrices between quarks and leptons appearing when we diagonalize all fermion masses following the following convention:
\begin{eqnarray}
&& u_{L} \to U_{L} u_{L}, \ d_{L} \to D_{L} d_{L}, \ e_{L} \to E_{L} e_{L}, \\
&& u_{R} \to U_{R} u_{R}, \ d_{R} \to D_{R} d_{R}, \ e_{R} \to E_{R} e_{R},
\end{eqnarray}
with
\begin{equation}
U_L^\dagger M_U U_R = M_U^{diag}, \ D_L^\dagger M_D D_R = M_D^{diag}, \ {\text{and}} \ E_L^\dagger M_E E_R = M_E^{diag}.  
\end{equation}
In Fig.~\ref{KLdecay} we show the vector leptoquark contributions to $K_L \to \mu^+ e^- $ decay. $K_L $ is written as
\begin{eqnarray}
    K_L = \frac{(1+\epsilon) K^0 + (1-\epsilon) \bar{K}^0}{\sqrt{2(1+|\epsilon|^2)}},
    \label{KLstate}
\end{eqnarray}
where $\epsilon \approx \mathcal{O}(10^{-3})$, $K^0 = (d\bar{s})$ and $\bar{K}^0 =( s\bar{d})$. Integrating out the vector leptoquark one finds the following effective Hamiltonian for $K_L^0 \to \mu^+ e^-$:
\begin{eqnarray}
{\mathcal{H}} &\supset& \frac{g_4^2}{2 M_X^2} \bar{d} \gamma^\nu (V_L^{d\mu} P_L + V_R^{d \mu} P_R) \mu \ \bar{e} 
\gamma_\nu ( (V_L^{se})^* P_L + (V_R^{se})^* P_R) s \nonumber \\
&+&  \frac{g_4^2}{2 M_X^2} \bar{s} \gamma^\nu (V_L^{s \mu} P_L + V_R^{s \mu} P_R) \mu \ \bar{e} \gamma_\nu ( (V_L^{de})^* P_L + (V_R^{de})^* P_R) d.
\end{eqnarray}
To rearrange the effective Hamiltonian for these decays, we use the following Fierz identities:
\begin{eqnarray}
\begin{split}
     & (\bar{\psi}_1 \gamma_\mu P_{L,R} \psi_2)\ (\bar{\psi}_3 \gamma^\mu P_{L,R} \psi_4)=(\bar{\psi}_1 \gamma_\mu P_{L,R} \psi_4) \ (\bar{\psi}_3 \gamma^\mu P_{L,R} \psi_2), \\
    & (\bar{\psi}_1 \gamma_\mu P_{L,R} \psi_2)(\bar{\psi}_3 \gamma^\mu P_{R,L} \psi_4)= -2  (\bar{\psi}_1 P_{R,L} \psi_4)(\bar{\psi}_3  P_{L,R} \psi_2).
\end{split}
\label{Fierz}
\end{eqnarray}
Therefore, for the $K_L \to \mu^+ e^-$ decay, the effective Hamiltonian reads as
\begin{equation}
\begin{split}
\mathcal{H} &\supset \left[ C_{V}^{(1)}(\bar{s}\gamma_\mu P_{L}d)+C_{V}^{(2)}(\bar{s}\gamma_\mu P_{R}d) + C_{V}^{(3)}(\bar{d}\gamma_\mu P_{L}s) + C_{V}^{(4)}(\bar{d}\gamma_\mu P_{R}s)\right]\hspace{0.1 cm}(\bar{e}\gamma^\mu\mu)\\
&+\left[C_{S}^{(1)}(\bar{s} P_{L}d) + C_{S}^{(2)}(\bar{s} P_{R}d)  + C_{S}^{(3)}(\bar{d} P_{L}s) +  C_{S}^{(4)}(\bar{d}P_{R}s) \right]\hspace{0.1 cm}(\bar{e}\mu) \\
& + \left[ C_{A}^{(1)}(\bar{s}\gamma_\mu P_{L}d)+C_{A}^{(2)}(\bar{s}\gamma_\mu P_{R}d) + C_{A}^{(3)}(\bar{d}\gamma_\mu P_{L}s) + C_{A}^{(4)}(\bar{d}\gamma_\mu P_{R}s)\right]\hspace{0.1 cm}(\bar{e}\gamma^\mu \gamma_5\mu)\\
&+\left[ C_{P}^{(1)}(\bar{s} P_{L}d)+C_{P}^{(2)}(\bar{s} P_{R}d) + C_{P}^{(3)}(\bar{d}P_{L}s) + C_{P}^{(4)}(\bar{d} P_{R}s)\right]\hspace{0.1 cm}(\bar{e} \gamma_5\mu), 
\end{split}
\label{h1}
\end{equation}
where the Wilson coefficients are given by
\begin{equation}
\begin{split}
&C_V^{(1)} \! = -\ C_{A}^{(1)} \! = \frac{g_4^2}{4M_X^2}(V_L^{de})^*V_L^{s\mu},  \quad
C_V^{(2)} \! = C_{A}^{(2)} \! = \frac{g_4^2}{4M_X^2}(V_R^{de})^*V_R^{s\mu}, \\
& C_V^{(3)} \! = -\  C_{A}^{(3)} \! = \frac{g_4^2}{4M_X^2}(V_L^{se})^*V_L^{d\mu}, \quad
C_V^{(4)} \! = \  C_{A}^{(4)} \! = \frac{g_4^2}{4M_X^2}(V_R^{se})^*V_R^{d\mu}, \\
& C_S^{(3)} \! = \  C_{P}^{(3)} \! =- \frac{g_4^2}{2M_X^2}(V_L^{se})^*V_R^{d\mu},\quad C_P^{(4)} \! = -\  C_{S}^{(4)} \! = \frac{g_4^2}{2M_X^2}(V_R^{se})^*V_L^{d\mu},\\
& C_S^{(1)} \! = \  C_{P}^{(1)} \! = - \frac{g_4^2}{2M_X^2}(V_L^{de})^*V_R^{s\mu},\quad C_P^{(2)} \! = -\  C_{S}^{(2)} \! = \frac{g_4^2}{2M_X^2}(V_R^{de})^*V_L^{s\mu}.
\end{split}
\end{equation}
To compute the amplitude for this process, the relevant matrix elements are given by
\begin{eqnarray}
\begin{split}
   & <0|\bar{d}\gamma_5s|K_L (p)>\ = \ <0|\bar{s}\gamma_5d|K_L(p)>\  = -i \frac{m_{K_L}^2}{m_d+m_s} \frac{F_K}{\sqrt{2}},
   \\
   &<0|\bar{d} \gamma^\nu\gamma_5s|K_L(p)>\ = \ <0|\bar{s} \gamma^\nu\gamma_5d|K_L(p)> \ = \ i \frac{F_K}{\sqrt{2}} \ p^\nu,
    \end{split}
\end{eqnarray}
with $F_K =155.72$ MeV and the Kaon mass, $m_{K_L}=497.6$ MeV. Here $p_\mu$ is the four-momentum of the Kaon meson.
Now, neglecting the electron mass and keeping only the non-zero matrix elements, the amplitude for this process reads 
\begin{equation}
    i{\cal M}= \frac{ i F_K}{2\sqrt{2}}\left[ C_V \ p_\nu \ (\bar{u}_e \gamma^\nu v_\mu )+ C_S \frac{m_{K_L}^2}{m_d +m_s}   \  (\bar{u}_e v_\mu) 
       + C_A \  p_\nu \ (\bar{u}_e \gamma^\nu \gamma_5 v_\mu )+ C_P \frac{m_{K_L}^2}{m_d + m_s} \  (\bar{u}_e \gamma_5v_\mu)\ \right],
\end{equation}
with the following combination of Wilson coefficients
\begin{eqnarray}
 C_V &=& C_V^{(2)}-C_V^{(1)} + C_{V}^{(4)}-C_{V}^{(3)}, \\
 C_S &=& C_S^{(1)}-C_S^{(2)} +C_S^{(3)}-C_S^{(4)}, 
 \\
 C_A &=& C_A^{(2)}-C_A^{(1)} +C_A^{(4)}-C_A^{(3)}, \\
 C_P &=& C_P^{(1)}-C_P^{(2)} +C_P^{(3)}-C_P^{(4)}.
\end{eqnarray}
Therefore, the decay width can be written as 
\begin{eqnarray}
    \Gamma(K_L^0 \to  \mu^+e^-)  &=&  \frac{F_K^2 \ m_{K_L}}{64 \pi}    \left(\left| C_S \ \frac{m_{K_L}^2}{m_d+m_s}   - m_\mu  \ C_V  \right|^2 +
\left| \ C_P \ \frac{m_{K_L}^2}{m_d +m_s}  + \ m_\mu \ C_A \right|^2 \  \right) \nonumber \\ 
&\times&
\ \left( 1-\frac{m_\mu^2}{m_{K_L}^2}\right)^2.
    \label{KLd1}
\end{eqnarray}
Finally, the total decay width is written as 
\begin{eqnarray}
\begin{split}
    \Gamma(K_L^0 \to \mu^\pm e^\mp) = \Gamma(K_L^0 \to \mu^-e^+)+\Gamma(K_L^0 \to \mu^+e^-) = 2\Gamma(K_L^0 \to \mu^+e^-).
    \end{split}
\end{eqnarray}
Notice that this theory does not predict the mixing matrices between quarks and leptons. Then, in general, one cannot predict the relevant Wilson coefficients for computing the Kaon decays. Therefore, the lower bound on the $SU(4)_C$ symmetry-breaking scale can be predicted only assuming some values for the mixing angles. 
As we show below, the lower bound can change dramatically when we assume different values for the $V_L$ and $V_R$ matrices.
It is important to mention that the Wilson coefficients appearing in the effective Hamiltonian are defined at the leptoquark mass ($M_{LQ}$) scale. However, in order to compare theoretical predictions with experimental results, these coefficients must be evolved down to the relevant low-energy scale using renormalization group running. As the electroweak correction is small compared to the QCD correction, neglecting the electroweak corrections, at 1 loop level, the coefficient evolves as~\cite{Buras:1998raa}
\begin{eqnarray}
    C(\mu) = U (\mu ,M_{LQ} ) \  C(M_{LQ}),
\end{eqnarray}
where 
\begin{equation}
    U(\mu ,M_{LQ}) = \left(\frac{g_s(\mu)}{g_s(m_c)}\right)^{8/\beta(n_f=3)} \left(\frac{g_s(m_c)}{g_s(m_b)}\right)^{8/\beta(n_f=4)} \left(\frac{g_s(m_b)}{g_s(m_t)}\right)^{8/\beta(n_f=5)}  \left(\frac{g_s(m_t)}{g_s(M_{LQ})}\right)^{8/\beta(n_f=6)},
\end{equation}
with $\beta(n_f)= 11 - (2/3) \ n_f$. Note that the vector and the axial currents do not run because of the Ward identity and, therefore, do not have an enhancement factor for the relevant Wilson coefficients. However, the scalar and pseudo-scalar operators would require a running factor to match the low-energy predictions.

\begin{figure}[t]
\includegraphics[width=0.55\textwidth]{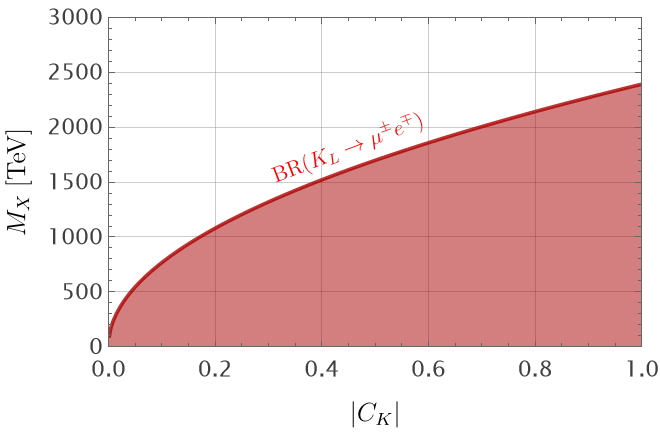}  
\caption{The red shaded region is excluded by the experiment bound ${\rm{BR}}(K_L \to e^\pm \mu^\mp) < 4.7 \times 10^{-12}$~\cite{BNL:1998apv}.}
\label{MXCK}
\end{figure}

One can simplify the decay width formula in Eq.(\ref{KLd1}) by assuming that the unknown mixing matrices are equal, $V_L =V_R =V$, and one finds
\begin{eqnarray}
    \Gamma(K_L\to \mu^\pm e^\mp)  = & \frac{1 }{2 M_X^4}  \pi \alpha_s^2\  F_K^2\  m_{K_L}  |C_K|^2  \left( 1-\frac{m_\mu^2}{m_{K_L}^2}\right)^2  \left(\frac{m_{K_L}^2}{m_d \ + \ m_s}-\frac{m_\mu}{2}\right)^2,
\end{eqnarray}
with only one effective Wilson coefficient
\begin{equation}
C_K= V_{de}^* V_{s\mu} + V_{se}^* V_{d\mu}.   
\end{equation}
In Fig.~\ref{MXCK} we show the lower bound on the vector-leptoquark mass as a function of the coefficient $C_K$. In the conservative scenario, with $V_{L}, V_R \sim 1$, the coefficient $C_K=1$ and one finds the lower bound: $M_X> 2.4 \times 10^5$ GeV. Here we use the current experimental value ${\rm{BR}}(K_L \to e^\pm \mu^\mp) < 4.7 \times 10^{-12}$~\cite{BNL:1998apv}.
However, in the general case, the value of $C_K$ is unknown, and the allowed value for $M_X$ could be close to the collider bounds. Then, one can hope to discover this theory at the Large Hadron Collider or any future collider.

As we have discussed above, in general we do not know the value of the mixing matrices $V_L$ and $V_R$, but we can consider some scenarios that can tell us how one can relax the most conservative lower bound on the symmetry breaking scale.
For simplicity, assuming that the mixing matrices $V_L, \ V_R$ are real, one can use the following parametrization:
\begin{eqnarray}
V_{{\rm{L,R}}} = V=
\begin{pmatrix}
c_{12} c_{13} &
s_{12} c_{13} &
s_{13} \\[6pt]

- s_{12} c_{23} - c_{12} s_{23} s_{13} &
\;\; c_{12} c_{23} - s_{12} s_{23} s_{13} &
s_{23} c_{13} \\[6pt]

s_{12} s_{23} - c_{12} c_{23} s_{13} &
- c_{12} s_{23} - s_{12} c_{23} s_{13} &
c_{23} c_{13}
\end{pmatrix},
\end{eqnarray}
where $s_{ij} = \sin{\theta_{ij}}$ and $c_{ij}=\cos{\theta_{ij}}$. Here the angles can be chosen between $0$ and $\pi/2$.

In order to discuss the impact of the unknown mixing angles between quarks and leptons on the value of the lower bound on the vector-leptoquark mass we study four simple scenarios:
\begin{itemize}
\item Case I: When $\theta_{13} = \theta_{23} =0$, the $V$ matrix reads as
\begin{eqnarray}
V =
\begin{pmatrix}
c_{12} & s_{12} & 0 \\
- s_{12} & c_{12} & 0 \\
0 & 0 & 1
\end{pmatrix}.
\end{eqnarray}
In this scenario, one can satisfy the experimental bounds from $K_L \to \mu^\pm e^\mp$  and $K_L \to e^+ e^-$ decays, if $M_X$ is above 1800 TeV. We show in Fig.~\ref{case1} the excluded colored regions using the experimental bounds. In red we show the excluded area by ${\rm{BR}}(K_L \to \mu^\pm e^\mp) < 4.7 \times 10^{-12}$~\cite{BNL:1998apv}, while in the light cyan, one shows the excluded region by $\Delta_K (e^+ e^-) < 5 \times 10^{-12}$~\cite{FileviezPerez:2008dw}.
\begin{figure}[h]
\centering
\includegraphics[width=0.55\textwidth]{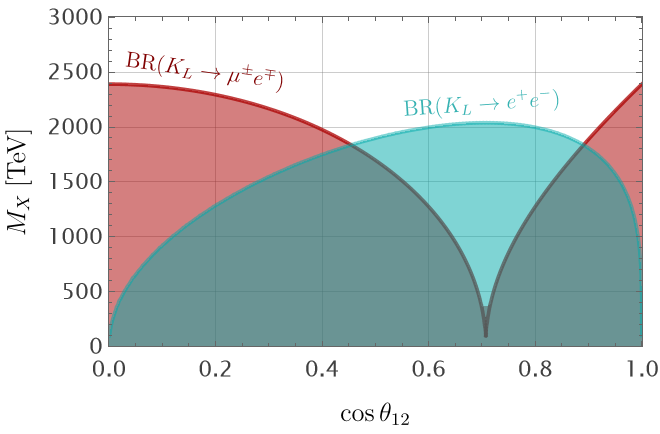}  
\caption{The shaded regions are excluded by the experimental bound ${\rm{BR}}(K_L \to \mu^\pm e^\mp) < 4.7 \times 10^{-12}$~\cite{BNL:1998apv} (region in red) and $\Delta_K (e^+ e^-) < 5 \times 10^{-12}$~\cite{FileviezPerez:2008dw} (region in light cyan).}
\label{case1}
\end{figure}
Notice that the bound from $K_L \to \mu^\pm e^\mp$ is weaker around $\cos \theta_{12} \sim 0.7$ because $C_K$ goes to zero. In the next section, we will show one could test this theory using the Mu2e experiment at Fermilab, even if the symmetry breaking scale is above $1.8 \times 10^3$ TeV.
\item Case II: In this scenario, $\theta_{13} = \pi/2 , \ \theta_{23} =0$, and the $V$ matrix is given by
\begin{eqnarray}
V =
\begin{pmatrix}
0 & 0 & 1 \\
- s_{12} & c_{12} & 0 \\
-c_{12} & -s_{12} & 0
\end{pmatrix}.
\end{eqnarray}
In this scenario, $\ \text{BR}(K_L \to \mu^\pm e^{\mp}) = \text{BR}(K_L \to e^+ e^{-})= \text{BR}(K_L \to \mu^+ \mu^{-}) =0$ at tree level. However, the $\mu \to e$ conversion experiments can provide strong bounds. We will discuss these bounds in the next section.
\item Case III: When $\theta_{13} = 0 , \ \theta_{23} =\pi/2$, the $V$ matrix becomes 
\begin{eqnarray}
V =
\begin{pmatrix}
c_{12} & s_{12} & 0 \\
0 & 0 & 1 \\
s_{12} & -c_{12} & 0
\end{pmatrix}.
\end{eqnarray}
In this scenario, $\ \text{BR}(K_L \to \mu^\pm e^{\mp}) = \text{BR}(K_L \to e^+ e^{-})= \text{BR}(K_L \to \mu^+ \mu^{-}) =0$ at tree level. As in Case II, $\mu \to e$ provides strong bounds in this scenario, see discussion in the next section.
\item Case IV: In this scenario, $\theta_{13} = \pi/2 , \ \theta_{23} =\pi/2$, the $V$ is given by
\begin{eqnarray}
V =
\begin{pmatrix}
0 & 0 & 1 \\
-c_{12} & -s_{12} & 0 \\
s_{12} & -c_{12} & 0
\end{pmatrix}.
\end{eqnarray}
As in the previous two cases, $\ \text{BR}(K_L \to \mu^\pm e^{\mp}) = \text{BR}(K_L \to e^+ e^{-})= \text{BR}(K_L \to \mu^+ \mu^{-}) =0$ at tree level. However, the $\mu \to e$ bounds discussed in the next section are quite important.
\end{itemize}
As one can appreciate, using the freedom in the unknown mixing angles between quarks and leptons, one can suppress or even avoid the strong constraints from the experimental bounds on the lepton flavour number violating $K_L$ decays. However, as we will discuss, the $\mu \to e$ bounds play a crucial role in constraining the $SU(4)_C$ breaking scale. This interplay between the meson decay bounds and $\mu \to e$, helps us to understand the possibility to test this theory in the near future. \vspace{-4pt}
\vspace{-4pt}
\begin{figure}[h]
\centering
\includegraphics[width=0.55\textwidth]{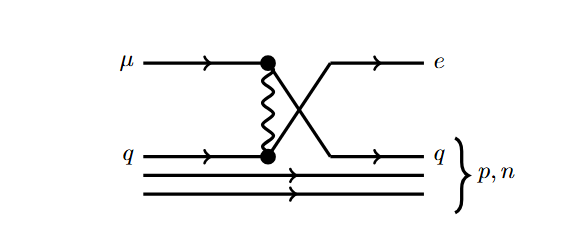}  
\caption{Feynman graph for $\mu \to e  $ conversion mediated by the vector leptoquark.}
\label{mutoeFeyn}
\end{figure}
%
\section {$\mu \to e $ CONVERSION}
\label{mutoe}
In Fig.~\ref{mutoeFeyn} we show the Feynman graph for $\mu \to e$ conversion mediated by the vector leptoquark in this theory.
After integrating out the vector leptoquark, the effective interacting Hamiltonian relevant for $\mu \to e$ conversion, can be written as 
\begin{equation}
\begin{split}
    \mathcal{H} \supset \  & C_{VL}^{(q)} \ (\bar{e}\gamma^\mu P_L\mu) \ (\bar{q}\gamma^\mu P_Lq  ) \ +  C_{VR}^{(q)} \ (\bar{e}\gamma^\mu P_R\mu) \ (\bar{q}\gamma^\mu P_R q  ) \\
    &  + C_{SR}^{(q)} \ (\bar{e} P_L\mu) \ (\bar{q} P_R  q  ) + C_{SL}^{(q)} \ (\bar{e} P_R\mu) \ (\bar{q} P_L  q  ).
\end{split}.
\label{Heff2}
\end{equation}
\begin{figure}[b]
    \centering
    \begin{subfigure}[t]{0.46\textwidth}
        \centering
        \includegraphics[width=\textwidth]{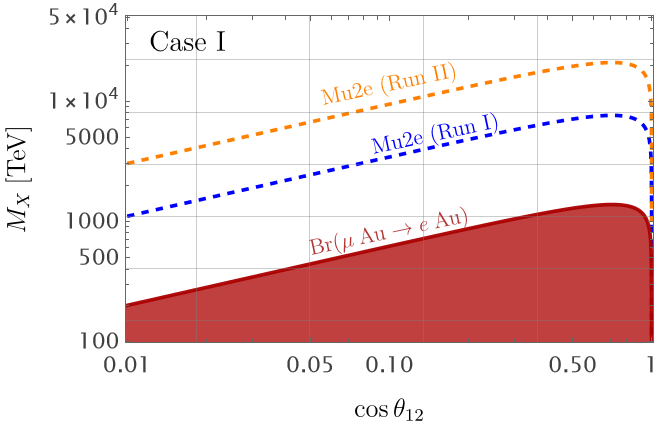}
        \caption{}
        \label{subfig:muecase1}
    \end{subfigure}  
    \begin{subfigure}[t]{0.45\textwidth}
        \centering
        \includegraphics[width=\textwidth]{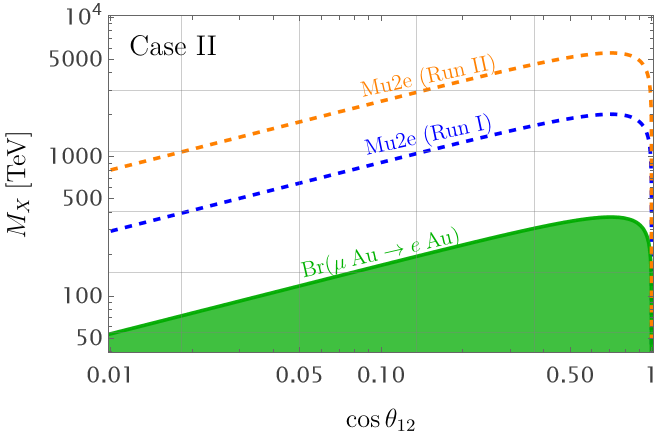}
        \caption{}
        \label{subfig:muecase2}
    \end{subfigure}          
    \begin{subfigure}[t]{0.46\textwidth}
        \centering
        \includegraphics[width=\textwidth]{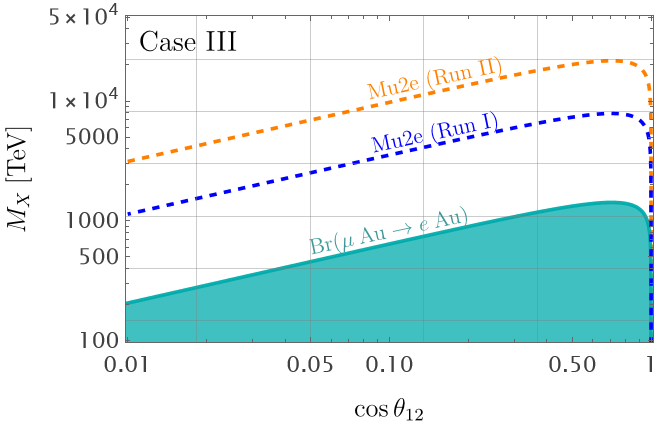}
        \caption{}
        \label{subfig:muecase3}
    \end{subfigure} 
    \begin{subfigure}[t]{0.45\textwidth}
        \centering
        \includegraphics[width=\textwidth]{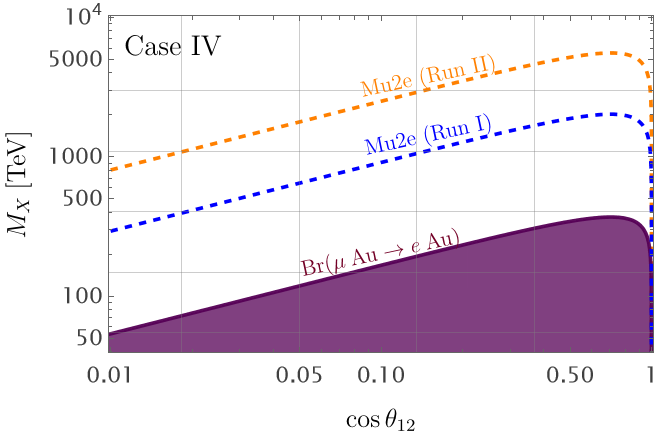}
        \caption{}
        \label{subfig:muecase4}
    \end{subfigure} 
    \caption{Predictions for $\mu  \to e$ conversion for different scenarios. In a) we show the predictions in Case I, where the colored region is excluded by the experimental bounds on $\mu \to e$ in Gold~\cite{SINDRUMII:2006dvw}. The blue and orange  dashed lines show the possible reach of the Mu2e experiment ~\cite{Mu2e:2014fns} for run I $( \text{BR}_{\mu \to e} \sim 6 \times 10^{-16} \ )$  and run II $( \text{BR}_{\mu \to e} \sim  10^{-17} \ )$, respectively. In b) we show the predictions for Case II, while the Case III predictions are shown in c). The predictions for Case IV are given in d). }
    \label{fig:mueconversion}
 \end{figure}
Here the Wilson coefficients for $ q= d,s $ are given by
\begin{equation}
\begin{split}
    & C_{VL}^{(q)} = \frac{g_4^2}{2 M_X^2} (V_L^{qe})^*\ V_L^{q\mu}, \quad  C_{VR}^{(q)} = \frac{g_4^2}{2 M_X^2} (V_R^{qe})^*\ V_R^{q\mu}, \\
    &  C_{SR}^{(q)} = - \frac{g_4^2}{ M_X^2} (V_R^{qe})^*\ V_L^{q\mu}, \quad  C_{S L}^{(q)} = -\frac{g_4^2}{ M_X^2} (V_L^{qe})^*\ V_R^{q\mu}.
\end{split}
\end{equation}
Notice that to derive the effective Hamiltonian in Eq.(\ref{Heff2}), we have used the Fierz identities in Eq. (\ref{Fierz}). Following Refs.~\cite{Kitano:2002mt,Crivellin:2017rmk,Cirigliano:2009bz}, the decay width for the coherent muon conversion in a nucleus can be written as
\begin{equation}
\begin{split}
     \Gamma_{\mu \to e} = &\frac{m_\mu^5}{16}\left[ \ \left| 4  \left(\tilde{C}_{VR \ }^{(p)}\ V^{(p)} +\tilde{C}_{VR}^{(n)} \ V^{(n)} + \tilde{C}^{(p)}_{SR}  \ S^{(p)} +\tilde{C}^{ (n)}_{SR} \ S^{(n)}  \right) \ \right|^2 \right. \\
    & + \left. \left| 4 \left(\tilde{C}_{VL}^{(p)} \ V^{(p)} +\tilde{C}_{VL }^{(n)} \ V^{(n)} +\tilde{C}^{ (p)}_{SL} \ S^{(p)} +\tilde{C}^{ (n)}_{SL} \ S^{(n)}  \right)\right|^2 \  \right],
\end{split}
\end{equation}
where we defined 
\begin{equation}
\begin{split}
        & \tilde{C}_{VR }^{(p)}= \sum_{q=u,d,s} C_{VR}^{(q)} \ f_{Vp}^{(q)} , \ \hspace{0.8 cm} \tilde{C}_{VL \ }^{(p)}= \sum_{q=u,d,s} C_{VL}^{(q)} \ f_{Vp}^{(q)}, \\
        & \tilde{C}_{VR }^{(n)}= \sum_{q=u,d,s} C_{VR}^{(q)} \ f_{Vn}^{(q)} , \ \hspace{0.8 cm} \tilde{C}_{VL \ }^{(n)}= \sum_{q=u,d,s} C_{VL}^{(q)} \ f_{Vn}^{(q)}, \\
        &\tilde{C}_{SR}^{(p)} = \sum_{q=u,d,s} \frac{m_p}{m_q} C_{SR}^{(q)} \ f_{Sp}^{(q)}, \hspace{0.4 cm}\tilde{C}_{SL}^{(p)} = \sum_{q=u,d,s} \frac{m_p}{m_q} C^{(q)}_{SL} \ f_{Sp}^{(q)},\\
        &\tilde{C}_{SR}^{(n)} = \sum_{q=u,d,s} \frac{m_N}{m_q} C_{SR}^{(q)} \ f_{Sn}^{(q)}, \hspace{0.4 cm} \tilde{C}_{SL}^{(n)} = \sum_{q=u,d,s} \frac{m_N}{m_q} C^{(q)}_{SL} \ f_{Sn}^{(q)}.
\end{split}
\end{equation}
The form factors for the vector operators are given by 
\begin{equation}
    \begin{split}
       &  f_{Vp}^{(d)} = 1  , \quad  f_{Vp}^{(s)} = 0 , \quad  f_{Vn}^{(d)} = 2  , \quad  f_{Vn}^{(s)} = 0, 
    \end{split}
\end{equation}
and the form factors for the scalar operators are given by~\cite{Hoferichter:2015dsa,Junnarkar:2013ac}
\begin{equation}
\begin{split}
    & f_{Sp}^{(u)} = (20.8 \pm 1.5 )\times 10^{-3}, \quad f_{Sp}^{(d)} = (41.1 \pm 2.8 )\times 10^{-3}, \quad f_{Sp}^{(s)} = (53 \pm 27)\times 10^{-3},\\
    & f_{Sn}^{(u)} = (18.9 \pm 1.4 )\times 10^{-3}, \quad f_{Sn}^{(d)} = (45.1 \pm 2.7 )\times 10^{-3}, \quad f_{Sn}^{(s)}. = (53 \pm 27)\times 10^{-3}.
\end{split}
\label{formfactor}
\end{equation}
Here, for convenience, we list the values of the overlap integrals and capture rates for muons for different nuclei in Table~\ref{tab:CoefficientChart}.
 \begin{table}[t]
\centering
\begin{tabular}{|c|c|c|c|c|c|}
\hline
\text{Nucleus} & \text{$S^p$} ~\cite{Kitano:2002mt} &\text{$S^n$} ~\cite{Kitano:2002mt} & \textbf{$V^p$} ~\cite{Kitano:2002mt} & \textbf{$V^n$} ~\cite{Kitano:2002mt} & \text{$\Gamma_{capt}^\mu (10^{6} s^{-1})$ ~\cite{Suzuki:1987jf}} \\
\hline
Al & 0.0155 & 0.0167 & 0.0161 & 0.0173 & 0.69 \\
\hline
Au & 0.0614 & 0.0918 & 0.0974 & 0.1460 & 13.07 \\
\hline
\end{tabular}
\caption{Values of the overlap integrals for different nuclei. }
\label{tab:CoefficientChart}
\end{table}
\begin{figure}[t]
 \centering
    \begin{subfigure}[b]{0.46\textwidth}
        \centering
        \includegraphics[width=\textwidth]{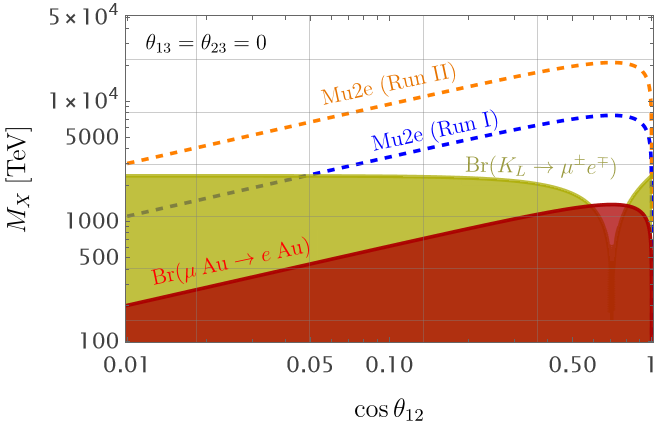}
        \caption{}
        \label{subfig:com11}
    \end{subfigure}  
    \begin{subfigure}[b]{0.45\textwidth}
        \centering
        \includegraphics[width=\textwidth]{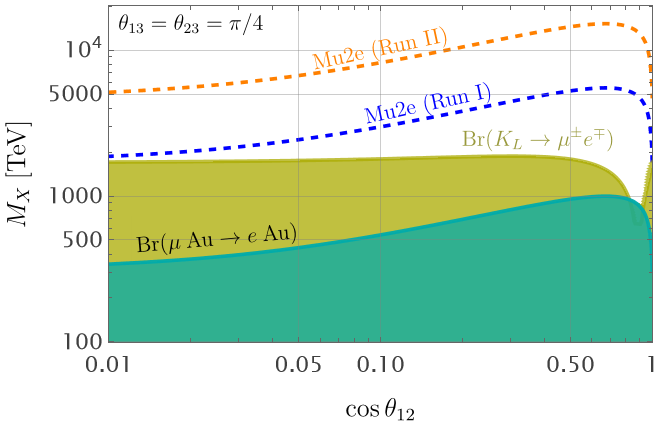}
        \caption{}
        \label{subfig:com2}
    \end{subfigure}     
\caption{Limits on the vector leptoquark mass. In a) one has the limits from $K_L$ decays and $\mu \to e$ when $\theta_{13}= \theta_{23}=0$. The red-shaded region is excluded by the $\mu \to e$ conversion in Gold~\cite{SINDRUMII:2006dvw}, and the rare Kaon decay excludes the yellow-shaded region. The dashed lines show the projected limits form the Mu2e experiment for $\mu \to e$ conversion in Al at Fermilab. b) Same as in a) but here we assume $\theta_{13}= \theta_{23}=\pi/4$. The cyan-shaded region is excluded by the $\mu \to e$ conversion in Gold~\cite{SINDRUMII:2006dvw}.}
\label{fig:ratio}
\end{figure}
In Fig.~\ref{fig:mueconversion} we show the predictions for $\mu  \to e$ conversion in the different scenarios discussed in the previous section. In Fig.~\ref{subfig:muecase1}, we show the predictions in Case I, where the colored region is excluded by the experimental bounds on $\mu \to e$ in Gold. In this case, one has the contribution of the $d$ and $s$ quarks to $\mu \to e$ and therefore the bounds are strong. In this case, the vector-leptoquark mass has be larger than approximately $10^3$ TeV when $\cos \theta_{12} \sim 1$. The expected exclusion regions by the Mu2e experiment are shown dashed lines. In blue (orange), one has the expected bound after Run I (II). Clearly, the Mu2e experiment can test the predictions for $\mu \to e$ in this theory or set stronger bounds on the $SU(4)_C$ symmetry breaking scale, $M_X \geq 10^4$ TeV. Notice that these bounds are much stronger than any collider bound for leptoquarks. 

In Fig.~\ref{subfig:muecase2} we show the predictions for Case II. In this scenario, only the $s$ quark contributes to the amplitude and one can expect weaker bounds. One can see in Fig.~\ref{subfig:muecase2} that the vector leptoquark mass can be much smaller. For example, $M_X \sim 60$ TeV when $\cos \theta_{12} \sim 0.01$. The Case III predictions are shown in Fig.~\ref{subfig:muecase3}. In this case one has only the $d$ quark contribution to $\mu \to e$, and then the bounds are similar to Case I. The predictions for Case IV are given in Fig.~\ref{subfig:muecase4}. This case is similar to Case II because one has only the $s$ quark contribution, and then the bounds obtained are much weaker. In any of these scenarios the Mu2e experiment could test the predictions or set very strong bounds. 

In Fig.~\ref{fig:ratio} we show the correlation between the current $\mu \to e$ constraints and the bounds from $K_L^0 \to e^\pm \mu^\mp$ decays.
For illustration, we show in Fig.~\ref{subfig:com11} the results when $\theta_{13}=\theta_{23}=0$, while in Fig.~\ref{subfig:com2} one assumes $\theta_{13}=\theta_{23}=\pi/4$. In both cases, one can see that currently the bounds from lepton number violating $K_L$ decays give us the strongest lower bound on the $SU(4)_C$ symmetry breaking scale. However, the projected Mu2e bounds shown by the dashed lines will exclude a larger region and can give us the most important bounds. Hopefully, the Mu2e can test the predictions for $\mu \to e$ conversion in this theory, providing a new path for physics beyond the Standard Model. 
\section{$\mu \to e$ CONVERSION AND SCALAR LEPTOQUARKS}
\label{scalarLQ}
In the minimal theory for quark-lepton unification discussed in this article, one can have extra contributions to lepton flavour violating processes. In Refs.~\cite{FileviezPerez:2021lkq,FileviezPerez:2022rbk} we studied some of the main LFV processes but not the constraints from $\mu \to e$ conversion.
In this theory, one can have extra constributions to $\mu \to e$ mediated by the new Higgses and scalar leptoquarks. However, only the scalar leptoquarks could have potential large contributions. 
The Yukawa interactions for the $\Phi_3$ scalar leptoquark can be written as
  \begin{equation}
    -{\cal L}  \supset  Y_2 \, \bar{\nu}_R \Phi_3^\alpha q_L^\beta \epsilon_{\alpha \beta} \ + Y_4 \, \bar{\ell}_L  \Phi_3 d_R +  {\rm H.c.}\, . 
  \end{equation}
  The $\Phi_3$ field can be written in $\SU(2)_L$ components as,
  \beq
  \Phi_3 =  \mqty(\phi_3^{1/3} \\[1ex]\phi_3^{-2/3} ),
  \eeq
  where the upper numbers denote the electric charge. The above interactions can be written explicitly as
  \begin{equation}
    -{\cal L}  \supset  Y_2 \, \bar{\nu}_R \phi_3^{1/3} d_L - Y_2 \bar{\nu}_R \phi_3^{-2/3} u_L \ + Y_4 \, \bar{\nu}_L  \phi_3^{1/3} d_R + Y_4 \bar{e}_L \phi_3^{-2/3} d_R \ + \ {\rm H.c.}\, . 
  \end{equation}
 Clearly, both leptoquarks in $\Phi_3$ can mediate $\mu \to e$ at tree level and the relevant coupling here is $Y_4$. As we have discussed in Ref.~\cite{Butterworth:2025ttw}, the Yukawa $Y_4$ is proportional to $M_D-M_E$, then generically these contributions are not large.
The $\Phi_4$ field has the following interactions with the SM fermions:
  \begin{equation}
    -{\cal L}  \supset Y_2 \, \bar{u}_R  \Phi_4^\alpha \ell_L^\beta \epsilon_{\alpha \beta} + \  Y_4 \, \bar{q}_L  \Phi_4 e_R +  {\rm H.c.}\, . 
  \end{equation}
  with
  \beq
  \Phi_4 =\mqty(\phi_4^{5/3} \\[1ex]\phi_4^{2/3} ).
  \eeq
One can write these Yukawa interactions explicitly as
  \begin{equation}
    -{\cal L}  \supset Y_2 \bar{u}_R \phi_4^{5/3} e_L - Y_2  \bar{u}_R \phi_4^{2/3} \nu_L  +  
    Y_4 \bar{u}_L \phi_4^{5/3} e_R + Y_4 \bar{d}_L \phi_4^{2/3} e_R + {\rm H.c.}.
  \end{equation}
\vspace{-7pt}
\begin{figure}[t]
\centering
\includegraphics[width=0.55\textwidth]{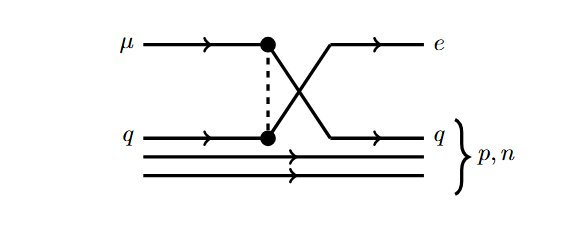}  
\caption{Feynman graph for $\mu \to e  $ conversion mediated by the scalar leptoquark.}
\label{mutoeFeynSc}
\end{figure}

As one can appreciate, the $\Phi_4^{5/3}$ can give us a contribution at tree level to $\mu \to e$ that is proportional to $Y_2$. The theory does not predict this Yukawa coupling, but in general, it could be large. Notice that the last term in the above equation is proportional to $Y_4$
and both $Y_2$ and $Y_4$  terms can contribute to the $\mu \to e$ conversion. However, $Y_4$ is proportional to the mass differences of the down quark and charged lepton, and it is naturally suppressed. Considering only the contribution from the $Y_2$ term, the effective Hamiltonian can be written as 
\begin{eqnarray}
    \mathcal{H} \supset \frac{1}{M_{\Phi_4}^2}y_2^{u\mu}{y_2^{ue}}^* (\bar{e} P_L u)\ (\bar{u} P_L \mu),
\end{eqnarray}
where $M_{\Phi_4^{}}$ is the mass of the  $\phi_4^{5/3}$ scalar leptoquark. Now, using the Fierz identity
\begin{eqnarray}
    (\bar{\psi}_1 P_L \psi_2)\ (\bar{\psi}_3 P_L \psi_4) = -\frac{1}{2} (\bar{\psi}_1 P_L \psi_4)(\bar{\psi}_3 P_L \psi_2) -\frac{1}{8} (\bar{\psi}_1\sigma_{\mu \nu} P_L \psi_4)(\bar{\psi}_3 \sigma^{\mu \nu} P_L \psi_2),
\end{eqnarray}
one can write the effective Hamiltonian as 
\begin{eqnarray}
    \mathcal{H} \supset \frac{1}{M_{\Phi_4}^2}y_2^{u\mu}{y_2^{ue}}^*\left(-\frac{1}{2}\  \bar{e} P_L \mu \ \bar{u} P_L u - \frac{1}{8} \ \bar{e} \sigma_{\mu \nu } P_L \mu \ \bar{u} \sigma^{\mu \nu } P_L u \right).
\end{eqnarray}
Here $y_2=U_R^\dagger Y_2 E_L$.
In the coherent $\mu \to e$ conversion the tensor operator doesn't contribute. Notice that in this scenario, only the up quark will contribute to this conversion process. Therefore, keeping only the scalar term, the conversion rate can be written as 
\begin{equation}
    \Gamma_{\mu \to e} = \frac{m_\mu^5}{4} \frac{|y_2^{u\mu}{y_2^{ue}}^*|^2}{M_{\Phi_4^{}}^4} \left| f_{Sp}^{(u)} \frac{m_p}{m_u} \ S^{(p)} + f_{Sn}^{(u)} \frac{m_N}{m_u} \ S^{(n)}\right|^2.
\end{equation}
The values of $f_{Sp}^{(u)}$ and $f_{Sn}^{(u)}$ are given in Eq.(\ref{formfactor}).
\begin{figure}[t]
\centering
\includegraphics[width=0.55\textwidth]{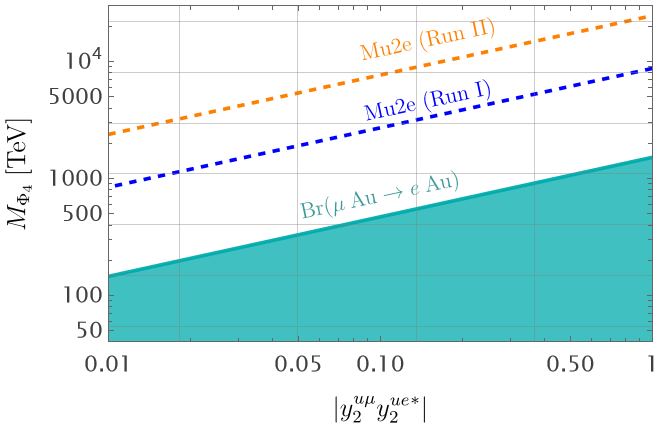}  
\caption{The colored region is excluded by the  $\mu \to e$ conversion in Au~\cite{SINDRUMII:2006dvw}, while the dashed lines show the projected bounds from the Mu2e experiment at Fermilab.}
\label{scalarMu2e}
\end{figure}
In Fig.~\ref{scalarMu2e} we show the excluded region by the current $\mu \to e$ conversion experiments when one considers only the contribution mediated by the $\Phi_4$ scalar leptoquark. Notice that when the $y_2$ Yukawa coupling is large, $y_2 \sim 1$, these bounds on the leptoquark mass, $M_{\Phi_4} > 10^3$ TeV, are much stronger than the collider bounds studied by us in a recent study in Ref.~\cite{Butterworth:2025ttw}. The projected bounds from the Mu2e experiment at Fermilab are shown by the blue and orange dashed lines. These bounds can exclude a large region of the parameter space and could reach the $M_{\Phi_4} \sim 10^4$ TeV when the Yukawa coupling is large. For the study of the $\mu \to e \gamma$ and $g-2$ contraints in this theory, see the study in Refs.~\cite{FileviezPerez:2021lkq,FileviezPerez:2022rbk}.
\section{SUMMARY}
\label{summary}
The idea of quark–lepton unification remains one of the most compelling frameworks for physics beyond the Standard Model. In this article, we have analyzed the predictions of the minimal theory of matter unification realized at the multi-TeV scale, where Majorana neutrino masses are generated through the inverse seesaw mechanism.
To assess the testability of this framework, we examined the impact of the most stringent experimental constraints and discussed the lower bound on the symmetry-breaking scale. A key prediction of this theory is the existence of both vector and scalar leptoquarks with masses in the multi-TeV range.

We have analyzed the predictions for lepton flavour violating meson decays in the minimal theory of quark–lepton unification, fully incorporating the general flavor structure and exploiting the freedom associated with the unknown quark–lepton mixing matrices for different chiralities.
Our results show that the symmetry-breaking scale can be much below $10^3$ TeV once the freedom on the mixing angles is taken into account. To illustrate this point, we present four simple benchmark scenarios that clearly identify the conditions under which the lepton flavor violating $K_L$ decays are strongly suppressed.

The predictions for $\mu$ to $e$ conversion have been investigated in detail, including the contributions mediated by both vector and scalar leptoquarks. We demonstrate that the Mu2e experiment at Fermilab can play a central role in testing the idea of quark–lepton unification.
We analyze the correlation between lepton flavor violating kaon decays and $\mu$ to $e$ conversion, showing that even in conservative scenarios where the matter unification scale lies above $10^3$ TeV, the projected sensitivity of Mu2e can probe symmetry breaking scales up to $10^4$ TeV.
These results indicate that the observation of $\mu$ to $e$ conversion at Fermilab would provide a powerful avenue to explore and potentially confirm the idea of matter unification in the near future.
\bibliography{refs}

@article{FileviezPerez:2022ypk,
    author = "Fileviez Perez, Pavel and others",
    title = "{On Baryon and Lepton Number Violation}",
    eprint = "2208.00010",
    archivePrefix = "arXiv",
    primaryClass = "hep-ph",
    month = "7",
    year = "2022"
}

@article{Babu:2025azv,
    author = "Babu, K. S. and Biswas, Sumit and Saad, Shaikh",
    title = "{TeV Scale Quark-Lepton Unification}",
    eprint = "2512.11113",
    archivePrefix = "arXiv",
    primaryClass = "hep-ph",
    month = "12",
    year = "2025"
}

@article{FileviezPerez:2008dw,
    author = "Fileviez Perez, Pavel and Han, Tao and Li, Tong and Ramsey-Musolf, Michael J.",
    title = "{Leptoquarks and Neutrino Masses at the LHC}",
    eprint = "0810.4138",
    archivePrefix = "arXiv",
    primaryClass = "hep-ph",
    reportNumber = "NPAC-08-19, MADPH-08-1522",
    doi = "10.1016/j.nuclphysb.2009.04.009",
    journal = "Nucl. Phys. B",
    volume = "819",
    pages = "139--176",
    year = "2009"
}

@article{Davidson:2022nnl,
    author = "Davidson, Sacha and Echenard, Bertrand",
    title = "{Reach and complementarity of $\mu \rightarrow e$ searches}",
    eprint = "2204.00564",
    archivePrefix = "arXiv",
    primaryClass = "hep-ph",
    doi = "10.1140/epjc/s10052-022-10773-4",
    journal = "Eur. Phys. J. C",
    volume = "82",
    number = "9",
    pages = "836",
    year = "2022"
}

@article{Davidson:2022jai,
    author = "Davidson, Sacha and Echenard, Bertrand and Bernstein, Robert H. and Heeck, Julian and Hitlin, David G.",
    title = "{Charged Lepton Flavor Violation}",
    eprint = "2209.00142",
    archivePrefix = "arXiv",
    primaryClass = "hep-ex",
    month = "8",
    year = "2022"
}

@article{Bernstein:2013hba,
    author = "Bernstein, Robert H. and Cooper, Peter S.",
    title = "{Charged Lepton Flavor Violation: An Experimenter's Guide}",
    eprint = "1307.5787",
    archivePrefix = "arXiv",
    primaryClass = "hep-ex",
    reportNumber = "FERMILAB-PUB-13-259-PPD",
    doi = "10.1016/j.physrep.2013.07.002",
    journal = "Phys. Rept.",
    volume = "532",
    pages = "27--64",
    year = "2013"
}

@article{FileviezPerez:2022rbk,
    author = "Fileviez Perez, Pavel and Murgui, Clara",
    title = "{Flavor anomalies and quark-lepton unification}",
    eprint = "2203.07381",
    archivePrefix = "arXiv",
    primaryClass = "hep-ph",
    doi = "10.1103/PhysRevD.106.035033",
    journal = "Phys. Rev. D",
    volume = "106",
    number = "3",
    pages = "035033",
    year = "2022"
}

@article{Valencia:1994cj,
    author = "Valencia, G. and Willenbrock, S.",
    title = "{Quark - lepton unification and rare meson decays}",
    eprint = "hep-ph/9409201",
    archivePrefix = "arXiv",
    reportNumber = "ILL-TH-94-17",
    doi = "10.1103/PhysRevD.50.6843",
    journal = "Phys. Rev. D",
    volume = "50",
    pages = "6843--6848",
    year = "1994"
}

@article{BNL:1998apv,
    author = "Ambrose, D. and others",
    collaboration = "BNL",
    title = "{New limit on muon and electron lepton number violation from K0(L) ---{\ensuremath{>}} mu+- e-+ decay}",
    eprint = "hep-ex/9811038",
    archivePrefix = "arXiv",
    doi = "10.1103/PhysRevLett.81.5734",
    journal = "Phys. Rev. Lett.",
    volume = "81",
    pages = "5734--5737",
    year = "1998"
}

@article{Butterworth:2025ttw,
    author = "Butterworth, Jon and Debnath, Hridoy and Fileviez Perez, Pavel and Wang, Peng",
    title = "{Quark-Lepton Unification Signatures}",
    eprint = "2512.00143",
    archivePrefix = "arXiv",
    primaryClass = "hep-ph",
    month = "11",
    year = "2025"
}

@article{Minkowski:1977sc,
    author = "Minkowski, Peter",
    title = "{$\mu \to e\gamma$ at a Rate of One Out of $10^{9}$ Muon Decays?}",
    reportNumber = "Print-77-0182 (BERN)",
    doi = "10.1016/0370-2693(77)90435-X",
    journal = "Phys. Lett. B",
    volume = "67",
    pages = "421--428",
    year = "1977"
}

@article{Gell-Mann:1979vob,
    author = "Gell-Mann, Murray and Ramond, Pierre and Slansky, Richard",
    title = "{Complex Spinors and Unified Theories}",
    eprint = "1306.4669",
    archivePrefix = "arXiv",
    primaryClass = "hep-th",
    reportNumber = "PRINT-80-0576",
    journal = "Conf. Proc. C",
    volume = "790927",
    pages = "315--321",
    year = "1979"
}

@article{Mohapatra:1979ia,
    author = "Mohapatra, Rabindra N. and Senjanovic, Goran",
    title = "{Neutrino Mass and Spontaneous Parity Nonconservation}",
    reportNumber = "MDDP-TR-80-060, MDDP-PP-80-105, CCNY-HEP-79-10",
    doi = "10.1103/PhysRevLett.44.912",
    journal = "Phys. Rev. Lett.",
    volume = "44",
    pages = "912",
    year = "1980"
}

@article{Yanagida:1979as,
    author = "Yanagida, Tsutomu",
    editor = "Sawada, Osamu and Sugamoto, Akio",
    title = "{Horizontal gauge symmetry and masses of neutrinos}",
    reportNumber = "KEK-79-18-95",
    journal = "Conf. Proc. C",
    volume = "7902131",
    pages = "95--99",
    year = "1979"
}

@article{Mohapatra:1986bd,
    author = "Mohapatra, R. N. and Valle, J. W. F.",
    title = "{Neutrino Mass and Baryon Number Nonconservation in Superstring Models}",
    reportNumber = "MdDP-PP-86-127",
    doi = "10.1103/PhysRevD.34.1642",
    journal = "Phys. Rev. D",
    volume = "34",
    pages = "1642",
    year = "1986"
}

@article{Smirnov:1995jq,
    author = "Smirnov, A. D.",
    title = "{The Minimal quark - lepton symmetry model and the limit on Z-prime mass}",
    eprint = "hep-ph/9503239",
    archivePrefix = "arXiv",
    reportNumber = "YARU-HE-95-01",
    doi = "10.1016/0370-2693(95)00015-D",
    journal = "Phys. Lett. B",
    volume = "346",
    pages = "297--302",
    year = "1995"
}

@article{Pati:1973rp,
    author = "Pati, Jogesh C. and Salam, Abdus",
    title = "{Is Baryon Number Conserved?}",
    reportNumber = "IC-73-85",
    doi = "10.1103/PhysRevLett.31.661",
    journal = "Phys. Rev. Lett.",
    volume = "31",
    pages = "661--664",
    year = "1973"
}

@article{Pati:1974yy,
    author = "Pati, Jogesh C. and Salam, Abdus",
    title = "{Lepton Number as the Fourth Color}",
    reportNumber = "IC-74-7",
    doi = "10.1103/PhysRevD.10.275",
    journal = "Phys. Rev. D",
    volume = "10",
    pages = "275--289",
    year = "1974",
    note = "[Erratum: Phys.Rev.D 11, 703--703 (1975)]"
}

@article{Gedeonova:2022iac,
    author = "Gedeonov{\'a}, Hedvika and Hudec, Mat{\v{e}}j",
    title = "{All possible first signals of gauge leptoquark in quark-lepton unification and beyond}",
    eprint = "2210.00347",
    archivePrefix = "arXiv",
    primaryClass = "hep-ph",
    doi = "10.1103/PhysRevD.107.095029",
    journal = "Phys. Rev. D",
    volume = "107",
    number = "9",
    pages = "095029",
    year = "2023"
}

@article{FileviezPerez:2013zmv,
    author = "Fileviez Perez, Pavel and Wise, Mark B.",
    title = "{Low Scale Quark-Lepton Unification}",
    eprint = "1307.6213",
    archivePrefix = "arXiv",
    primaryClass = "hep-ph",
    doi = "10.1103/PhysRevD.88.057703",
    journal = "Phys. Rev. D",
    volume = "88",
    pages = "057703",
    year = "2013"
}

@article{FileviezPerez:2022fni,
    author = "Fileviez Perez, Pavel and Golias, Elliot and Plascencia, Alexis D.",
    title = "{Two-Higgs-doublet model and quark-lepton unification}",
    eprint = "2205.02235",
    archivePrefix = "arXiv",
    primaryClass = "hep-ph",
    doi = "10.1007/JHEP08(2022)293",
    journal = "JHEP",
    volume = "08",
    pages = "293",
    year = "2022"
}

@article{FileviezPerez:2021arx,
    author = "Fileviez Perez, Pavel and Golias, Elliot and Plascencia, Alexis D.",
    title = "{Probing quark-lepton unification with leptoquark and Higgs boson decays}",
    eprint = "2107.06895",
    archivePrefix = "arXiv",
    primaryClass = "hep-ph",
    doi = "10.1103/PhysRevD.105.075011",
    journal = "Phys. Rev. D",
    volume = "105",
    number = "7",
    pages = "075011",
    year = "2022"
}

@article{FileviezPerez:2021lkq,
    author = "Fileviez Perez, Pavel and Murgui, Clara and Plascencia, Alexis D.",
    title = "{Leptoquarks and matter unification: Flavor anomalies and the muon g-2}",
    eprint = "2104.11229",
    archivePrefix = "arXiv",
    primaryClass = "hep-ph",
    doi = "10.1103/PhysRevD.104.035041",
    journal = "Phys. Rev. D",
    volume = "104",
    number = "3",
    pages = "035041",
    year = "2021"
}

@article{Flacke:2025xwl,
    author = "Flacke, Thomas and Kim, Jeong Han and Kunkel, Manuel and Pi, Jun Seung and Porod, Werner",
    title = "{Hunting and identifying coloured resonances in four top events with machine learning}",
    eprint = "2506.04318",
    archivePrefix = "arXiv",
    primaryClass = "hep-ph",
    reportNumber = "KIAS - A25019",
    month = "6",
    year = "2025"
}

@article{FileviezPerez:2023rxn,
    author = "Fileviez P\'erez, Pavel and Murgui, Clara and Patrone, Samuel and Testa, Adriano and Wise, Mark B.",
    title = "{Finite naturalness and quark-lepton unification}",
    eprint = "2308.07367",
    archivePrefix = "arXiv",
    primaryClass = "hep-ph",
    reportNumber = "Report-no: CALT-TH/2023-025",
    doi = "10.1103/PhysRevD.109.015011",
    journal = "Phys. Rev. D",
    volume = "109",
    number = "1",
    pages = "015011",
    year = "2024"
}

@article{Murgui:2021bdy,
    author = "Murgui, Clara and Wise, Mark B.",
    title = "{Scalar leptoquarks, baryon number violation, and Pati-Salam symmetry}",
    eprint = "2105.14029",
    archivePrefix = "arXiv",
    primaryClass = "hep-ph",
    doi = "10.1103/PhysRevD.104.035017",
    journal = "Phys. Rev. D",
    volume = "104",
    number = "3",
    pages = "035017",
    year = "2021"
}

@article{ATLAS:2023uox,
    author = "Aad, Georges and others",
    collaboration = "ATLAS",
    title = "{Search for pair production of third-generation leptoquarks decaying into a bottom quark and a $\tau $-lepton with the ATLAS detector}",
    eprint = "2303.01294",
    archivePrefix = "arXiv",
    primaryClass = "hep-ex",
    reportNumber = "CERN-EP-2022-267",
    doi = "10.1140/epjc/s10052-023-12104-7",
    journal = "Eur. Phys. J. C",
    volume = "83",
    number = "11",
    pages = "1075",
    year = "2023"
}

@article{Hoferichter:2015dsa,
    author = "Hoferichter, Martin and Ruiz de Elvira, J. and Kubis, Bastian and Mei{\ss}ner, Ulf-G.",
    title = "{High-Precision Determination of the Pion-Nucleon {\ensuremath{\sigma}} Term from Roy-Steiner Equations}",
    eprint = "1506.04142",
    archivePrefix = "arXiv",
    primaryClass = "hep-ph",
    reportNumber = "INT-PUB-15-026",
    doi = "10.1103/PhysRevLett.115.092301",
    journal = "Phys. Rev. Lett.",
    volume = "115",
    pages = "092301",
    year = "2015"
}

@article{Junnarkar:2013ac,
    author = "Junnarkar, Parikshit and Walker-Loud, Andre",
    title = "{Scalar strange content of the nucleon from lattice QCD}",
    eprint = "1301.1114",
    archivePrefix = "arXiv",
    primaryClass = "hep-lat",
    reportNumber = "NT-LBNL-13-001, UCB-NPAT-13-001, UNH-13-01",
    doi = "10.1103/PhysRevD.87.114510",
    journal = "Phys. Rev. D",
    volume = "87",
    pages = "114510",
    year = "2013"
}

@article{Kitano:2002mt,
    author = "Kitano, Ryuichiro and Koike, Masafumi and Okada, Yasuhiro",
    title = "{Detailed calculation of lepton flavor violating muon electron conversion rate for various nuclei}",
    eprint = "hep-ph/0203110",
    archivePrefix = "arXiv",
    reportNumber = "KEK-TH-808",
    doi = "10.1103/PhysRevD.76.059902",
    journal = "Phys. Rev. D",
    volume = "66",
    pages = "096002",
    year = "2002",
    note = "[Erratum: Phys.Rev.D 76, 059902 (2007)]"
}

@article{Crivellin:2017rmk,
    author = "Crivellin, Andreas and Davidson, Sacha and Pruna, Giovanni Marco and Signer, Adrian",
    title = "{Renormalisation-group improved analysis of $\mu\to e$ processes in a systematic effective-field-theory approach}",
    eprint = "1702.03020",
    archivePrefix = "arXiv",
    primaryClass = "hep-ph",
    reportNumber = "PSI-PR-17-01, ZU-TH-01-17",
    doi = "10.1007/JHEP05(2017)117",
    journal = "JHEP",
    volume = "05",
    pages = "117",
    year = "2017"
}

@article{Suzuki:1987jf,
    author = "Suzuki, T. and Measday, David F. and Roalsvig, J. P.",
    title = "{Total Nuclear Capture Rates for Negative Muons}",
    reportNumber = "TRI-PP-87-5",
    doi = "10.1103/PhysRevC.35.2212",
    journal = "Phys. Rev. C",
    volume = "35",
    pages = "2212",
    year = "1987"
}

@article{CMS:2025iix,
    author = "Hayrapetyan, A. and others",
    collaboration = "CMS",
    title = "{Search for $t$-channel scalar and vector leptoquark exchange in the high-mass dimuon and dielectron spectra in proton-proton collisions at $\sqrt{s}$ = 13 TeV}",
    eprint = "2503.20023",
    archivePrefix = "arXiv",
    primaryClass = "hep-ex",
    reportNumber = "CMS-EXO-22-013, CERN-EP-2024-275",
    doi = "10.1007/JHEP12(2025)052",
    journal = "JHEP",
    volume = "12",
    pages = "052",
    year = "2025"
}

@article{CMS:2022goy,
    author = "Tumasyan, Armen and others",
    collaboration = "CMS",
    title = "{Searches for additional Higgs bosons and for vector leptoquarks in $\tau\tau$ final states in proton-proton collisions at $\sqrt{s}$ = 13 TeV}",
    eprint = "2208.02717",
    archivePrefix = "arXiv",
    primaryClass = "hep-ex",
    reportNumber = "CMS-HIG-21-001, CERN-EP-2022-137",
    doi = "10.1007/JHEP07(2023)073",
    journal = "JHEP",
    volume = "07",
    pages = "073",
    year = "2023"
}

@article{CMS:2024bnj,
    author = "Hayrapetyan, Aram and others",
    collaboration = "CMS",
    title = "{Search for pair production of scalar and vector leptoquarks decaying to muons and bottom quarks in proton-proton collisions at s=13{\,}{\,}TeV}",
    eprint = "2402.08668",
    archivePrefix = "arXiv",
    primaryClass = "hep-ex",
    reportNumber = "CMS-EXO-21-019, CERN-EP-2023-301",
    doi = "10.1103/PhysRevD.109.112003",
    journal = "Phys. Rev. D",
    volume = "109",
    number = "11",
    pages = "112003",
    year = "2024"
}

@article{Cirigliano:2009bz,
    author = "Cirigliano, Vincenzo and Kitano, Ryuichiro and Okada, Yasuhiro and Tuzon, Paula",
    title = "{On the model discriminating power of mu ---{\ensuremath{>}} e conversion in nuclei}",
    eprint = "0904.0957",
    archivePrefix = "arXiv",
    primaryClass = "hep-ph",
    reportNumber = "IFIC-09-14, FTUV-09-0402, KEK-TH-1308, LA-UR-09-01473, TU-845",
    doi = "10.1103/PhysRevD.80.013002",
    journal = "Phys. Rev. D",
    volume = "80",
    pages = "013002",
    year = "2009"
}

@article{SINDRUMII:2006dvw,
    author = "Bertl, Wilhelm H. and others",
    collaboration = "SINDRUM II",
    title = "{A Search for muon to electron conversion in muonic gold}",
    doi = "10.1140/epjc/s2006-02582-x",
    journal = "Eur. Phys. J. C",
    volume = "47",
    pages = "337--346",
    year = "2006"
}

@article{Mu2e:2014fns,
    author = "Bartoszek, L. and others",
    collaboration = "Mu2e",
    title = "{Mu2e Technical Design Report}",
    eprint = "1501.05241",
    archivePrefix = "arXiv",
    primaryClass = "physics.ins-det",
    reportNumber = "FERMILAB-TM-2594, FERMILAB-DESIGN-2014-01",
    doi = "10.2172/1172555",
    month = "10",
    year = "2014"
}

@inproceedings{Buras:1998raa,
    author = "Buras, Andrzej J.",
    title = "{Weak Hamiltonian, CP violation and rare decays}",
    booktitle = "{Les Houches Summer School in Theoretical Physics, Session 68: Probing the Standard Model of Particle Interactions}",
    eprint = "hep-ph/9806471",
    archivePrefix = "arXiv",
    reportNumber = "TUM-HEP-316-98",
    pages = "281--539",
    month = "6",
    year = "1998"
}
\end{document}